\documentclass[aps,prb,twocolumn,showpacs,preprintnumbers,amsmath,amssymb,superscriptaddress]{revtex4}

\usepackage{graphicx}
\usepackage{dcolumn}
\usepackage{bm}
\usepackage{color}

\newcommand{\ket}[1]{|#1\rangle}

\begin{document}

\title{Negative tunnel magnetoresistance and differential conductance in transport
through double quantum dots}

\author{Piotr Trocha}
\email{piotrtroch@gmail.com}\affiliation{Department of Physics,
Adam Mickiewicz University, 61-614 Pozna\'n, Poland}

\author{Ireneusz Weymann}
\affiliation{Department of Physics, Adam Mickiewicz University,
61-614 Pozna\'n, Poland} \affiliation{Department of Physics,
Arnold Sommerfeld Center for Theoretical Physics, Ludwig
Maximilians Universit\"at M\"unchen, Theresienstr. 37, 80333
Munich, Germany}

\author{J\'ozef Barna\'s}
\affiliation{Department of Physics, Adam
Mickiewicz University, 61-614 Pozna\'n, Poland}
\affiliation{Institute of Molecular Physics, Polish Academy of
Sciences, 60-179 Pozna\'n, Poland}

\date{\today}

\begin{abstract}
Spin-dependent transport through two coupled single-level quantum
dots weakly connected to ferromagnetic leads with collinear
magnetizations is considered theoretically. Transport
characteristics, including the current, linear and nonlinear
conductance, and tunnel magnetoresistance are calculated using the
real-time diagrammatic technique in the parallel, serial, and
intermediate geometries. The effects due to virtual tunneling
processes between the two dots {\it via}  the leads, associated
with off-diagonal coupling matrix elements, are also considered.
Negative differential conductance and negative tunnel
magnetoresistance have been found in the case of serial and
intermediate geometries, while no such behavior has been observed
for double quantum dots coupled in parallel. It is  also shown
that transport characteristics strongly depend on the magnitude of
the off-diagonal coupling matrix elements.
\end{abstract}

\pacs{72.25.-b, 73.23.Hk, 73.63.Kv, 85.75.-d}

\maketitle


\section{Introduction}


Transport properties of double quantum dots (DQDs) have recently
attracted considerable attention from both experimental and
theoretical sides. \cite{derWielRMP03,lossPRA98,hansonPRL07,
ono02,graberPRB06,mcclurePRL07,golovachPRB04,
cotaPRL05,aghassiPRB06,wunschPRB05,franssonPRB06} This is mainly
due to the fact that DQDs are one of the simplest model systems
that mimic  behavior of real molecules, and are thus frequently
referred to as {\it artificial molecules}. Moreover, double
quantum dots are considered to play an important role in quantum
computation \cite{lossPRA98,hansonPRL07} and spintronics.
\cite{maekawa02,zutic04,maekawa06} They exhibit a variety of
different phenomena, including the Pauli spin blockade,
\cite{ono02} formation of molecular states, \cite{graberPRB06}
spin filtering effects, \cite{cotaPRL05} or various interference
effects, such as Fano or Dicke resonances. \cite{trocha,trochaJNN}
In addition, very recently it was shown theoretically that, when
coupled to ferromagnetic leads, double quantum dots display a
considerable tunnel magnetoresistance (TMR) and spin accumulation
effects. \cite{hornbergerPRB08,weymannPRB_DQDs,trocha,trocha1}
Spin-dependent transport properties of quantum dots have been so
far mainly addressed in the case of single quantum dots.
\cite{rudzinski01,braunPRB04,cottet04,weymannPRB05,kondoSQD,barnasJPCM08}
This field is already rather well established and transport
through single quantum dots coupled to ferromagnetic leads has
been extensively studied experimentally.
\cite{sahoo05,pasupathy04,fertAPL06, hamayaAPL07a, hamayaAPL07b,
hamayaPRB08,parkinNL08,hamayaPRL09} On the other hand, theoretical
investigations of spin effects in multi-dot structures are in
relatively initial stage, and so is experimental implementation of
DQDs coupled to ferromagnetic leads, which still remains a
challenge.

In this paper we consider the spin-dependent transport properties
of double quantum dots focusing on the weak coupling regime.
Conductance of the system is then determined mainly by
discreteness of the dots' energy levels and Coulomb correlations,
which may lead to the Coulomb blockade effect and step-like
current-voltage characteristics. To calculate the transport
characteristics in the linear and nonlinear response regimes, we
employ the real-time diagrammatic technique. \cite{diagrams} This
technique allows us to take into account the interference effects
resulting from virtual processes between the two quantum dots and
the leads as well as renormalization of the dot levels. In
particular, taking into account the first-order self-energies, we
calculate the current, conductance and tunnel magnetoresistance
for various geometries of the double quantum dot system. In
particular, we analyze the transport characteristics in the case
of DQDs connected in series, in parallel, as well as for some
intermediate geometries. We show that  the interference effects
associated with off-diagonal matrix elements of the self-energy
can significantly influence transport properties of the system for
parallel and intermediate geometries. When the quantum dots are
coupled in series or are in an intermediate geometry, we find
negative differential conductance (NDC) and negative TMR in some
transport regimes. These features appear in transport through DQD
systems. However, they were not found in transport through a
single quantum dot connected to ferromagnetic leads in the
corresponding range of parameters. Furthermore, we also analyze
the dependence of transport properties on the magnitude of the
off-diagonal matrix elements. Finally, we note that in previous
theoretical considerations,
\cite{wunschPRB05,aghassiPRB06,weymannPRB_DQDs} virtual
first-order tunneling processes have not been taken into account
as they become relevant for parallel and intermediate geometries
of double quantum dot systems.

The paper is organized as follows. In Sec. 2 we describe the model
of a double quantum dot and outline the method used in
calculations. Numerical results on the current, conductance and
tunnel magnetoresistance for DQDs coupled in serial, in parallel
and for intermediate geometries are presented and discussed in
Sec. 3. The main focus here is on negative differential
conductance and negative tunnel magnetoresistance. Summary and
final conclusions are given in Sec. 4.


\section{Theoretical description}


\subsection{Model}

\begin{figure}[t]
  \includegraphics[width=0.65\columnwidth]{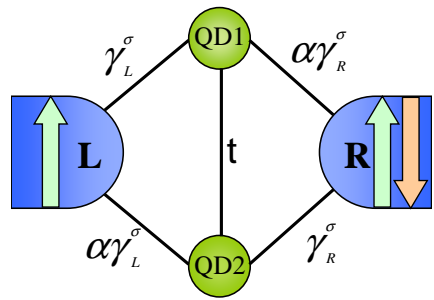}
  \caption{\label{Fig:1}
  (Color online) Schematic picture of the DQD system
   coupled to ferromagnetic leads. The parameter $\gamma_{\beta}^\sigma$
  (for $\beta=L, R, \sigma=\uparrow ,\downarrow$) describes here a
  spin-dependent dot-lead couplings, whereas $\alpha$ takes into
  account difference in the coupling of a given electrode to the two
  dots ($\alpha\in\langle 0,1\rangle$).
  In particular, for $\alpha=0$ double quantum dots are in the serial geometry,
  while for $\alpha=1$ the system is in the parallel geometry.
  }
\end{figure}

We consider two coupled single-level quantum dots connected to
ferromagnetic leads, as shown schematically in Fig.~\ref{Fig:1}.
The magnetizations of the leads are assumed to be collinear, and
the system can be either in the parallel or antiparallel magnetic
configuration. The system can be switched from one configuration
to the other by applying a weak external magnetic field and
sweeping through the hysteresis loop, provided the leads have
different coercive fields. The Hamiltonian of the double quantum
dot system is generally given by
\begin{equation} \label{Eq:Hamiltonian}
  H = H_{\rm leads} + H_{\rm DQD} + H_{\rm tunnel},
\end{equation}
where the first term, $H_{\rm leads}$, describes the left (L) and
right (R) electrodes in the non-interacting quasi-particle
approximation, $H_{\rm leads} = H_{\rm L} + H_{\rm R}$, with
$H_\beta = \sum_{\mathbf{k}\sigma}
\varepsilon_{\beta\mathbf{k}\sigma}
c^{\dagger}_{\beta\mathbf{k}\sigma} c_{\beta\mathbf{k}\sigma}$
(for $\beta={\rm L,R}$). Here,
$c^{\dagger}_{\beta\mathbf{k}\sigma}$
($c_{\beta\mathbf{k}\sigma}$) is the creation (annihilation)
operator of an electron with the wave vector $\mathbf{k}$ and spin
$\sigma$ in the lead $\beta$, whereas
$\varepsilon_{\beta\mathbf{k}\sigma}$ denotes the corresponding
single-particle energy. The second term of the Hamiltonian
describes the double quantum dot and is given by,
\begin{eqnarray}
   H_{\rm DQD}&=&\sum_{i\sigma}\limits\varepsilon_{i\sigma}d^\dag_{i\sigma}d_{i\sigma}
   + \sum_i\limits
   U_in_{i\sigma}n_{i\bar{\sigma}}
   \nonumber \\
    &+&U_0(n_{1\uparrow}+n_{1\downarrow})(n_{2\uparrow}+n_{2\downarrow}),
\end{eqnarray}
where $\bar{\sigma}\equiv -\sigma$, $n_{i\sigma} =
d^\dag_{i\sigma}d_{i\sigma}$ is the particle number operator for
spin $\sigma$ in the dot $i$ ($i=1,2$), $d^\dag_{i\sigma}$
($d_{i\sigma}$) is the respective creation (annihilation)
operator, and $\varepsilon_{i\sigma}$ denotes the spin-dependent
discrete energy level of the $i$-th dot. Double occupation of the
dot $i$ is associated with the intradot charging energy $U_i$,
whereas simultaneous occupation of both dots with one electron per
dot costs the interdot charging energy $U_0$. In the following we
assume $U_1=U_2\equiv U$, and note that $U_0<U$ for typical
lateral double dot structures. \cite{derWielRMP03} We further
parameterize the quantum dot energy levels by their average
energy, $E_\sigma = (\varepsilon_{1\sigma} +
\varepsilon_{2\sigma})/2$ and their difference, $\Delta E =
\varepsilon_{1\sigma} - \varepsilon_{2\sigma}$, respectively, so
that $\varepsilon_{1\sigma}=E_{\sigma}+\Delta E/2$ and
$\varepsilon_{2\sigma}=E_{\sigma}-\Delta E/2$. Here we assumed
that the dots have equal $g$-factors, then $\Delta E$ is
independent of spin even in the presence of external magnetic
field. Furthermore, we also assume $E_\uparrow=E_\downarrow\equiv
E$, if not stated otherwise. Apart from this, we assume that the
bare energy levels of the dots are independent of the applied
transport voltage. This can be achieved for instance with suitable
gate voltages.

The last term of the Hamiltonian, Eq.~(\ref{Eq:Hamiltonian}),
consists of two different terms, $H_{\rm tunnel}=H_{\rm V}+H_{\rm
t}$. The first one describes the spin-dependent tunneling
processes between the quantum dots and external magnetic leads and
is given by
\begin{equation}
   H_{\rm V} = \sum_{\beta i}\sum_{\mathbf{k}\sigma}
    \left( V_{\beta i} c^\dag_{\beta\mathbf{k}\sigma}d_{i\sigma}
    + V_{\beta i}^* d_{i\sigma}^\dagger c_{\beta\mathbf{k}\sigma}
    \right),
\end{equation}
where $V_{\beta i}$ are the relevant tunneling matrix elements
between the lead $\beta$ and dot $i$. The second term of $H_{\rm
tunnel}$ corresponds to hopping between the two quantum dots and
reads
\begin{equation}
   H_{\rm t} = -t\sum_\sigma\left(d^\dag_{1\sigma}d_{2\sigma}
             + d^\dag_{2\sigma}d_{1\sigma} \right).
\end{equation}
The inter-dot hopping parameter $t$ is assumed to be real and
independent of the electron spin orientation. We also assume that
all tunneling processes in the system are spin-conserving.

Due to the coupling to external leads, the dot levels acquire
finite widths. The dot-leads coupling is described generally by
$\Gamma_{\beta ij}^\sigma = 2\pi \rho_\beta^\sigma V_{\beta i}
V_{\beta j}^\ast$, where $\rho_{\beta}^\sigma$ is the density of
states of the lead $\beta$ for spin $\sigma$, $\sigma=+(-)$ for
the majority (minority) spin electrons. $\Gamma_{\beta ij}^\sigma$
describes the spin-dependent hybridization between the local dot
levels ($i,j=1,2$) and the leads, and is directly related to the
coupling strength between the dots and leads. In principle, the
coupling parameters may be energy-dependent. However, for
transport regimes considered in this paper, it is well justified
to assume that the couplings are constant within the electron
band. \cite{wunschPRB05} For the considered system, the coupling
parameters can be conveniently written in a matrix form as
\begin{equation}\label{Eq:coupling}
  \mathbf{\Gamma}_\beta^\sigma = \left(
\begin{array}{cc}
  \Gamma^\sigma_{\beta 11} & \Gamma^\sigma_{\beta 12} \\
  \Gamma^\sigma_{\beta 21} & \Gamma^\sigma_{\beta 22}
\end{array}
\right),
\end{equation}
where the tunneling matrix elements $V_{\beta i}$ are assumed to
be real and constant, while $\Gamma^\sigma_{\beta 12} =
\Gamma^\sigma_{\beta 21} = q_\beta (\Gamma^\sigma_{\beta 11}
\Gamma^\sigma_{\beta 22})^{1/2}$. The off-diagonal matrix elements
of $\mathbf{\Gamma}_\beta^\sigma$ are associated with various
interference effects resulting from virtual first-order tunneling
processes between the two quantum dots through the states in the
leads. These off-diagonal matrix elements may be significantly
reduced in comparison to diagonal matrix elements
$\Gamma^\sigma_{\beta ii}$. Furthermore, for complete destructive
interference these matrix elements may be totally suppressed. To
take the interference effects into account, we have introduced the
parameters $q_{\rm L}$ and $q_{\rm R}$. (For calculation of the
parameters $q_L$ and $q_R$ see Ref.~[\onlinecite{kubo}].) We
further assumed that $q_\beta$ are real positive numbers and
fulfill the condition $q_\beta \le 1$. Moreover, by introducing
the spin polarization of lead $\beta$, $p_\beta = (\rho_\beta^+
-\rho_\beta^- )/ (\rho_\beta^+ +\rho_\beta^- )$, the coupling
constants in the parallel configuration can be simply written as
\begin{equation}\label{Eq:coupling_left}
  \mathbf{\Gamma}_{\rm L}^{\uparrow(\downarrow)} = \gamma_L^{\sigma}\left(
\begin{array}{cc}
  1 & q_{\rm L}\sqrt{\alpha} \\
  q_{\rm L}\sqrt{\alpha} & \alpha
\end{array}
\right),
\end{equation}
for the coupling to the left electrode and
\begin{equation}\label{Eq:coupling_right}
  \mathbf{\Gamma}_{\rm R}^{\uparrow(\downarrow)} = \gamma_R^{\sigma}\left(
\begin{array}{cc}
  \alpha & q_{\rm R}\sqrt{\alpha} \\
  q_{\rm R}\sqrt{\alpha} & 1
\end{array}
\right),
\end{equation}
for coupling to the right lead. In the above expressions
$\gamma_L^{\sigma}=(1\pm p_{\rm L})\Gamma_L$ and
$\gamma_R^{\sigma}=(1\pm p_{\rm R})\Gamma_R$.  Here, we assume
that the couplings are symmetric, $\Gamma_{\rm L} = \Gamma_{\rm
R}\equiv \Gamma/2$, and $\alpha$ takes into account the difference
in the coupling of a given electrode to the dots, see
Fig.~\ref{Fig:1}. In principle, the parameter $\alpha$ could be
different for the left and right leads. However,  we assume here
that the system is symmetric, as shown in Fig.1. In the above
formulas we have also assumed that in the parallel configuration
the spin-$\uparrow$ (spin-$\downarrow$) electrons belong to the
majority (minority) electron bands of the leads. In the
antiparallel configuration the couplings are given by
Eqs.~(\ref{Eq:coupling_left}) and (\ref{Eq:coupling_right}) with
$p_{\rm R} \leftrightarrow - p_{\rm R}$. By varying the parameter
$\alpha$, one can change the geometry of the system from serial
one for $\alpha = 0$ to the parallel geometry for $\alpha = 1$.
For intermediate values of $\alpha$ the system is in an
intermediate geometry, where each of the two dots is coupled to
both leads, see Fig.~\ref{Fig:1}. It is worth noting that
investigating the effect of geometry on transport properties by
varying the parameter $\alpha$ is certainly relevant from
experimental point of view.

\subsection{Method}

In order to determine the transport properties of the system we
employ the real-time diagrammatic technique. \cite{diagrams} This
technique is based on the perturbation expansion of the reduced
density matrix and the relevant operators with respect to the
coupling strength $\Gamma$. We calculate the reduced density
matrix $\hat{\rho}$ for the double-dot system by integrating out
the electronic degrees of freedom in the leads. The time evolution
of $\hat{\rho}$ is then described by the Liouville equation of the
form \cite{diagrams,wunschPRB05}
\begin{equation} \label{Eq:Liouville}
  i\hbar\frac{d}{dt}\mathbf{\hat{\rho}} =
    [H_{\rm DQD}+H_{\rm t},\mathbf{\hat{\rho}}] +
    \Sigma\mathbf{\hat{\rho}}\;.
\end{equation}
The commutator represents the internal dynamics in the double dot,
which mainly depends on the level separation $\Delta E$ and the
interdot coupling $t$. The second part of Eq.~(\ref{Eq:Liouville})
accounts for the tunnel coupling between the double dot and
external reservoirs. The complex tensor $\Sigma$ is associated
with tunneling processes and tunnel-induced energy renormalization
of the dot levels. The elements of the reduced density matrix are
defined as
$P^{\chi_1}_{\chi_2}\equiv\langle\chi_1|\mathbf{\hat{\rho}}|\chi_2\rangle$,
where $\chi_1$ and $\chi_2$ denote the eigenstates of the DQD
system. Then, the Liouville equation for stationary reduced
density matrix can be written in the form \cite{diagrams}
\begin{equation} \label{Eq:probability}
  0 = \langle\chi_1|[H_{\rm DQD} + H_{\rm t},\mathbf{\hat{\rho}}]\chi_2\rangle
  + \sum_{\chi_1' \chi_2'} \Sigma_{\chi_2 \chi_2'}^{\chi_1 \chi_1'}
  P^{\chi_1'}_{\chi_2'} \;.
\end{equation}
Here, $\Sigma_{\chi_2\chi_2'}^{\chi_1\chi_1'}$ denotes the
self-energy corresponding to evolution forward in time from state
$\ket{\chi_1'}$ to state $\ket{\chi_1}$ and then backward in time
from state $\ket{\chi_2}$ to state $\ket{\chi_2'}$. The diagonal
elements of the reduced density matrix, $P^{\chi}_{\chi}$
($\chi_1=\chi_2=\chi$), correspond to probability of finding the
DQD system in the state $\ket{\chi}$. To solve Eq.9 for density
matrix elements one usually performs a perturbation expansion with
respect to the dot-lead coupling strength $\Gamma$. Then, each
term of the expansion can be visualized graphically as a diagram
defined on the Keldysh contour, where the vertices are connect by
lines corresponding to tunneling processes. The self-energies in
respective order of expansion can be calculated using the
diagrammatic rules.~\cite{diagrams}

In our considerations we take into account the limit of weak
tunnel coupling between the dots. For serial geometry of the
double dot system, tunneling between the two dots becomes then a
bottle-neck for transport and may considerably alter the
spin-dependent transport characteristics of the system, as
presented in the next section. This is contrary to previous
theoretical studies of spin-dependent transport in
DQDs,~\cite{weymannPRB_DQDs} where the hopping between the two
dots was relatively large and transport took place through highly
hybridized molecular-like states of the system. To determine
transport characteristics in the weak coupling regime we perform
systematic perturbation expansion with respect to the coupling
parameter $\Gamma$. Furthermore, we assume $\Gamma \ll k_{B}T$.
The current is then mediated mainly by first-order (sequential)
tunneling processes, while the higher-order coherent tunneling
events play a minor role, and it is justifiable to neglect
them.~\cite{wunschPRB05,aghassiPRB06} We thus investigate the
basic transport properties using the sequential tunneling
approximation, i.e. we need to determine only the lowest-order
self-energies which involve one tunneling line. Some examples of
first-order diagrams relevant for the present calculation are
shown in the Appendix.

After calculating the density matrix elements from
Eq.~(\ref{Eq:probability}), one can determine the sequential
current flowing through the double dot system from the following
formula
\begin{equation}\label{Eq:current}
    I = -\frac{ie}{2\hbar}\sum_{\substack{\chi_1\chi_2 \\
    \chi_1'\chi_2'}}
    {\Sigma^{\rm I}}_{\chi_2\chi_2'}^{\chi_1\chi_1'} P_{\chi_2'}^{\chi_1'},
\end{equation}
where ${\Sigma^{\rm I}}_{\chi_2\chi_2'}^{\chi_1\chi_1'}$ denotes
the first-order self-energy in which one vertex was substituted by
a vertex representing the current operator,
$\hat{I}=(\hat{I}_R-\hat{I}_L)/2$, with
$\hat{I}_\beta=-i(e/\hbar)\sum_{i}\sum_{\mathbf{k}\sigma}
    ( V_{\beta i} c^\dag_{\beta\mathbf{k}\sigma}d_{i\sigma}
    - V_{\beta i}^* d_{i\sigma}^\dagger
    c_{\beta\mathbf{k}\sigma})$.

In our analysis we assume that the intra-dot charging energy $U$
is relatively large for both dots, much larger than the interdot
Coulomb correlation energy $U_0$. Thus, only the zero, one and
two-particle DQD states are relevant for transport. Furthermore,
in the case of $U \gg U_0$, the occupation probability of the
states with two electrons in the same dot is also vanishingly
small. However, these states are taken into account as
intermediate (virtual) states in the calculation, providing a
natural high-energy cutoff. Finally, we would like to emphasize
that by using the real-time diagrammatic technique we are able to
take into account the first-order virtual tunneling processes
between the two dots in a fully systematic
way.~\cite{diagrams,wunschPRB05}


\section{Numerical results}


In this section we present and discuss numerical results on the
charge current, differential conductance and tunnel
magnetoresistance of double quantum dots coupled in serial,
parallel, and intermediate geometries. The TMR effect results
generally from spin-dependent dot-lead tunneling processes, which
in turn leads to the dependence of transport characteristics on
magnetic configuration of the system. The TMR is quantitatively
described by the ratio ${\rm TMR} = (I_{\rm P}-I_{\rm AP})/I_{\rm
AP}$, where $I_{\rm P}$ and $I_{\rm AP}$ denote the currents
flowing through the system in the parallel and antiparallel
magnetic configurations, respectively.~\cite{julliere75,barnas98}

In the numerical analysis we assume spin degenerate dot levels,
$\varepsilon_{i\sigma} = \varepsilon_i$ (for $i =1,2$ and $\sigma
=\uparrow ,\downarrow$). We also assume that external electrodes
are made of the same ferromagnetic material, $p_L=p_R\equiv p$,
and that the system is symmetrically coupled to the leads,
$\Gamma_{\rm L}=\Gamma_{\rm R}\equiv\Gamma/2$. The parameters
$q_{\rm L}$ and $q_{\rm R}$ can be generally different. However,
we assume that they are real and equal, $q_{\rm L} = q_{\rm R}
\equiv q$. We also set the intra- and inter-dot Coulomb parameters
to be: $U=100k_BT$ and $U_0=20k_BT$, respectively. Finally, we
assume spin polarization $p=0.4$, which is typical of 3d
ferromagnetic metals. \cite{Soulen98} The inter-dot hopping
parameter $t$ is assumed to be: $t=0.25\Gamma$ with $\Gamma=5\mu
$eV. These are typical experimental parameters for double quantum
dot systems. \cite{ono02,derWielRMP03} Chemical potentials of the
left and right leads are set to be $\mu_L=eV/2$ and $\mu_R=-eV/2$,
where $eV=\mu_L-\mu_R$ denotes the applied bias voltage.

\begin{figure}[t]
  \includegraphics[width=0.8\columnwidth]{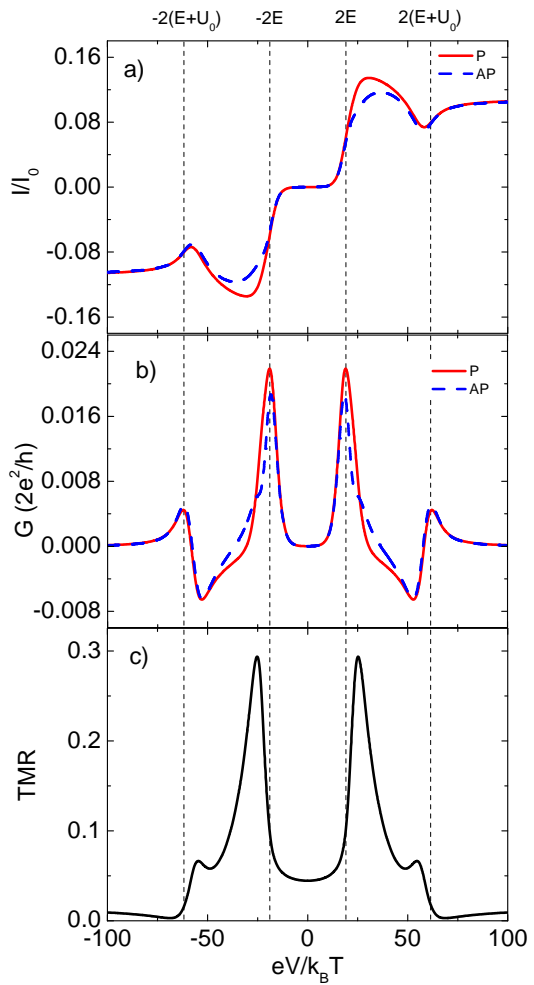}
  \caption{\label{Fig:2}
  (Color online) Current (a) and differential conductance (b) in the parallel (P, solid line)
  and antiparallel (AP, dashed line) magnetic configurations, as well as tunnel
  magnetoresistance (c), calculated as a function of the bias
  voltage   for the parameters:
  $E=10 k_BT$, $\Delta E=0$, $U_0=20 k_BT$, $U=100 k_BT$, $p=0.4$,
  $t=0.25\Gamma$, $\Gamma=5\mu$eV, $\alpha=0$, and
  $I_0=e\Gamma/\hbar\approx 1.215$ nA.
  }
\end{figure}

\subsection{Double dots connected in series, $\alpha =0$}

\begin{figure}[t]
  \includegraphics[width=0.8\columnwidth]{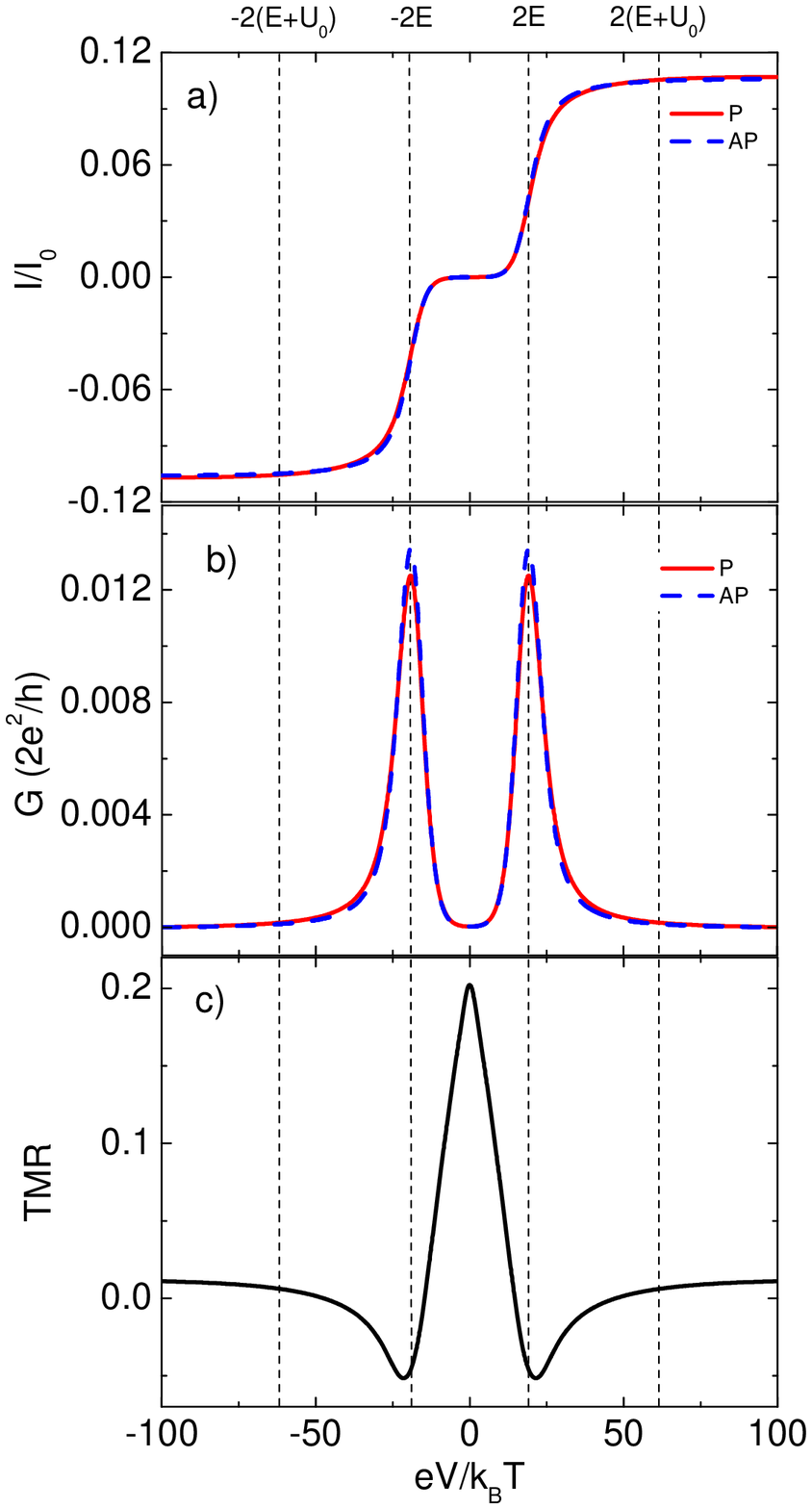}
  \caption{\label{Fig:3}
  (Color online) Current (a) and differential conductance (b) in the parallel (solid line)
  and antiparallel (dashed line) magnetic configurations, and tunnel
  magnetoresistance (c), calculated as a function of the bias voltage for $E=-10 k_BT$.
  The other parameters are the same as in Fig.~\ref{Fig:2}.}
\end{figure}

Let us first consider the situation when $\alpha=0$, which
corresponds to serial geometry of the double quantum dot system,
see Fig.~\ref{Fig:1}. In Fig.~\ref{Fig:2} we show the basic
transport characteristics for the average dot level $E=10 k_BT$
and the difference between bare dots' levels $\Delta E =0$. In the
weak coupling regime transport is determined mainly by
discreteness of the energy dot spectrum and Coulomb correlations,
which lead to staircase-like current-voltage characteristics, see
Fig.~\ref{Fig:2}(a). For the assumed parameters, the DQD is empty
at low bias  and the current is blocked below the threshold
voltage, irrespective of magnetic configuration of the system. In
the blockade regime, however, the first-order processes can still
contribute to the current due to thermal fluctuations.
Furthermore, in the case of $\Gamma \ll k_{B}T$, as considered in
this paper, the contribution from first-order processes can still
be larger than that from second-order tunneling (cotunneling).
Nevertheless, one must bear in mind that in the case of {\em deep}
Coulomb blockade, for instance for $E/k_BT \ll 0$ and
$(E+U_0)/k_BT\gg 0$, the second-order processes become dominant
and must be taken into account to properly describe transport
properties of the system.~\cite{weymannPRB_DQDs} In this paper,
however, we restrict ourselves to the case when transport is
mainly governed by sequential tunneling processes.

When the bias voltage approaches the threshold voltage, the
sequential current starts to flow due to one-by-one electron
tunneling through singly-occupied DQD states. This leads to the
peak in the differential conductance, see Fig.~\ref{Fig:2}(b).
However, when the bias voltage increases further, instead of a
plateau one observes a drop of the current, which leads to
negative differential conductance. This feature appears in both
magnetic configurations, see Fig.~\ref{Fig:2}(b). When $eV$
approaches $2E+2U_0$, where another electron has possibility to
tunnel into the DQD system, the current starts increasing further.

Physical mechanism responsible for the occurrence of negative
differential conductance follows from the level renormalization
due to tunneling processes between the dots and leads. This
renormalization is directly related to the real part of the
off-diagonal self-energies, see Eq.~(\ref{diagram2}) (and also
Ref.~[\onlinecite{wunschPRB05}], where the level renormalization
in a DQD system connected in series and coupled to nonmagnetic
leads was calculated). Accordingly, the renormalized level
$\varepsilon_i^{\rm ren,\sigma}$ of the $i$th dot for spin
$\sigma$ has the following form:
\begin{eqnarray}\label{Eq:renor}
  \varepsilon_i^{\rm ren,\sigma}&=&\varepsilon_i+\Omega_{\alpha
  i}^{\sigma}(E) +\Omega_{\alpha i}^{\sigma}(E+U) \nonumber\\
  &-& \Omega_{\alpha i}^{\uparrow}(E+U_0) - \Omega_{\alpha
  i}^{\downarrow}(E+U_0)\,,
\end{eqnarray}
with $\Omega_{\beta i}^{\sigma}(x) = (\Gamma_{\beta ii} ^ {\sigma}
/ 2\pi) \left[{\rm Re} \Psi \left( \frac{1}{2} +
i\frac{x-\mu_\beta}{2\pi k_BT} \right)\right]$, where $\Psi(x)$ is
the digamma function. This renormalization lifts the initially
assumed degeneracy of the two dot's levels. The larger separation
between these renormalized levels, the smaller probability of
electron tunneling from the left to the right dot. We have
calculated this separation as a function of the bias voltage (not
shown) and found that the separation increases with increasing
bias voltage (in the voltage range where negative differential
conductance appears), and therefore the current decreases with
increasing bias. After reaching maximum, the level separation
starts to decrease with a further increase in voltage and negative
differential conductance disappears. It is also worth noting that,
due to the coupling to ferromagnetic leads, the level
renormalization becomes spin dependent and, consequently, depends
on magnetic configuration of the system, and so does the level
spacing.

The above described  level renormalization makes the occupation
probability of the left dot (QD1) for $eV>E$ larger than the
occupation probability of the right dot (QD2), $P_{|\sigma
,0\rangle}>P_{|0,\sigma\rangle}$ for $\sigma=\uparrow ,
\downarrow$. Moreover, the rate for tunneling between the two dots
decreases with increasing the bias voltage. More specifically,
${\rm Im}\langle 0,\sigma|\mathbf{\hat{\rho}}|\sigma ,0\rangle$
decreases, whereas ${\rm Im}\langle \sigma,
0|\mathbf{\hat{\rho}}|0,\sigma \rangle$ increases, $\langle\sigma
,0|\mathbf{\hat{\rho}}|0,\sigma\rangle=(\langle
0,\sigma|\mathbf{\hat{\rho}}|\sigma ,0\rangle)^{\ast}$. In the
case of tunneling through double quantum dots connected in series,
the imaginary parts of the density matrix, e.g. $\langle
0,\sigma|\mathbf{\hat{\rho}}|\sigma ,0\rangle$ and $\langle
\sigma, 0|\mathbf{\hat{\rho}}|0,\sigma \rangle$, are related to
the charge transfer through the system, i.e. they are directly
related to the current flow. \cite{wunschPRB05} Consequently, the
tunneling of electrons from the left lead to the left quantum dot
(QD1) is partially suppressed because of $P_{|\sigma
,0\rangle}>P_{|0,\sigma\rangle}$ and due to decreased tunneling
rates between the two dots with increasing bias voltage $V$ in the
range $|eV|\in (2E, 2E+2U_0)$.

The tunnel magnetoresistance as a function of applied bias voltage
is plotted in Fig.~\ref{Fig:2}(c). Since the conductance is larger
in the parallel configuration than in the antiparallel one, the
corresponding TMR is positive, although very small in the voltage
range where double occupancy is allowed. Moreover, one can note
that the sequential tunneling TMR is generally smaller than the
Julliere's value of TMR, \cite{julliere75}
$\rm{TMR}^{Jull}=2p^2/(1-p^2)\approx 0.38$ for $p=0.4$, which is
characteristic of single tunnel junction or fully coherent
transport. \cite{weymannPRB05} In the case shown in
Fig.~\ref{Fig:2}(c), the TMR reaches local maxima for $|eV|=2E$
and for $|eV|=2E+2U_0$, but it is especially enhanced at the first
Coulomb step, i.e. in the vicinity of $|eV|=2E$.

In Fig. \ref{Fig:3} we present results obtained for DQD connected
in series with average double dot level position being negative
$E=-10k_BT$. The DQD system is then singly occupied in
equilibrium, and the system is in the Coulomb blockade regime as
double occupation of DQD would cost the correlation energy $U_0$.
Since the levels corresponding to $E$ and $E+U_0$ start
contributing to current at the same bias ($E$ and $E+U_0$ are
symmetric with respect to the zero bias voltage), only one step is
observed in the bias dependence of current. Contrary to the case
presented in Fig.~\ref{Fig:2}, now the current is a monotonic
function of the applied bias voltage, and negative differential
conductance is not observed, see Fig.~\ref{Fig:3}(a) and (b). The
difference between the two magnetic configurations is only
slightly visible in the current and differential conductance.
Accordingly, the TMR  above the threshold voltage is very small,
see Fig.~\ref{Fig:3}(c). At the zero bias, however, TMR exhibits a
sharp maximum and then drops with increasing bias voltage.
Furthermore, for voltages around the resonance $|eV|\approx 2E$ a
negative TMR is observed.

\begin{figure}[t]
  \includegraphics[width=0.8\columnwidth]{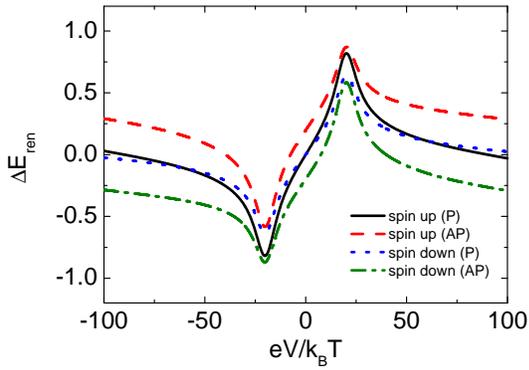}
  \caption{\label{Fig:4}
  (Color online) Renormalization of the dots' levels spacing for both spin
  orientations in the parallel and antiparallel magnetic configurations.
  The parameters are the same as in Fig.~\ref{Fig:3}.}
\end{figure}

\begin{figure}[t]
  \includegraphics[width=0.9\columnwidth]{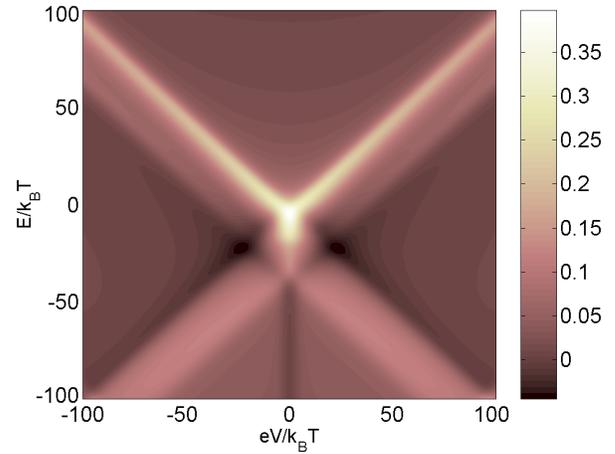}
  \caption{\label{Fig:5}
  (Color online) Tunnel magnetoresistance as a function of the bias
  voltage and the average level position. The other
   parameters as in Fig.~\ref{Fig:3}.}
\end{figure}

Physical origin of the negative TMR is similar to that of negative
differential conductance discussed above. More specifically,
negative TMR follows from the level renormalization due to
coupling of the dots to external leads in the presence of
inter-dot Coulomb correlations. This level renormalization is spin
dependent, so that it lifts spin degeneracy of the dot levels.
More importantly, it is generally different for each dot and
therefore modifies the renormalized level spacing $\Delta E_{\rm
ren,\sigma}=\varepsilon_{1}^{\rm ren,\sigma}-\varepsilon_2^{\rm
ren,\sigma}$. The renormalization of the level spacing in the
system under consideration is displayed in Fig.~\ref{Fig:4}. For
the assumed parameters hopping between the dots is like a {\it
bottle neck} for electrons, that controls current flowing through
the system. As already stated above, this hopping probability
decreases with increasing level separation (for each spin
orientation). When comparing Fig.~\ref{Fig:3}(c) and
Fig.~\ref{Fig:4}, one can note that the minimum in (negative) TMR
appears in the  voltage range where the level spacings are
maximum. Let us consider in more details positive bias, $eV>0$. In
the region where negative TMR is observed, the level spacing
between left and right dots for the dominant transport channel
(spin-up) in the parallel magnetic configuration is significantly
larger than that for spin-down electrons in the parallel
configuration and also significantly larger than the spacing for
one of the spin channels in the antiparallel configuration, while
it is comparable to the level spacing for second spin channel in
the antiparallel configuration. Thus, in the parallel
configuration the spin-down channel, which involves spin-minority
bands in both leads, takes over control of the current, while the
dominant spin channel in the antiparallel configuration involves
one spin-majority and one spin-minority bands. Consequently, the
current is then larger in the antiparallel configuration than in
the parallel one, which results in negative TMR.

To support this, we have calculated the relevant occupation
probabilities. We have found that the occupation probability of
the left dot by a spin-up electron in the parallel configuration
is larger than that for the antiparallel configuration
configuration, $P^P_{L\uparrow}>P^{AP}_{L\uparrow}$, whereas
opposite relation holds for spin-down electrons,
$P^P_{L\downarrow}<P^{AP}_{L\downarrow}$. This situation is
opposite to that found for the bias region where large positive
TMR appears. Moreover, the occupation probabilities of the right
dot by a spin-up or spin-down electron in both magnetic
configurations are small in comparison with those for the left
dot. Finally, the probability of double occupation of the DQD
system is rather small, consequently such states are rather
irrelevant.

\begin{figure}[t]
  \includegraphics[width=0.8\columnwidth]{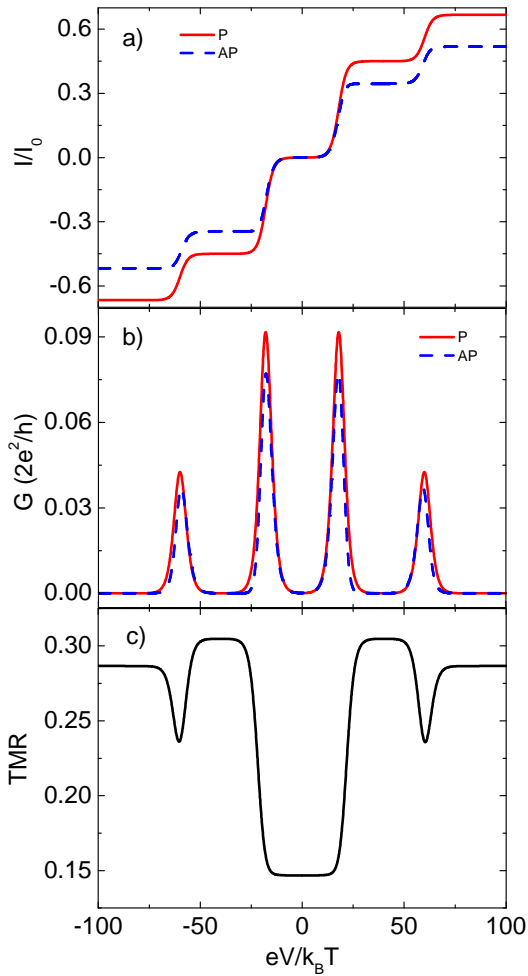}
  \caption{\label{Fig:6}
  (Color online) The current (a), differential conductance (b) in the parallel (solid line)
  and antiparallel (dashed line) configurations, and tunnel
  magnetoresistance (c) as a function of the bias voltage obtained
  for the parameters: $\alpha=1$, $q=0.25$, $E=10 k_BT$, while the other parameters
  are as in Fig.~\ref{Fig:2}.}
\end{figure}

In Fig.~\ref{Fig:5} we show the bias voltage and level position
dependence of the TMR. The position of the average level can be
changed experimentally by sweeping the gate voltage, so that
Fig.~\ref{Fig:5} effectively shows the bias and gate voltage
dependence of TMR. One can note that as the absolute value of the
average level position increases, the central maximum in TMR
becomes split into two components, while the maximum at zero bias
changes into a minimum. In the case of negative $E$, the minimum
at zero bias occurs roughly when $E\lesssim -U_0$. Furthermore,
there are also transport regions where TMR changes sign and
becomes negative. The negative TMR occurs mainly for $E\approx
-U_0/2$ and $|eV|\approx 2E$. One can also note that generally TMR
becomes much suppressed for larger bias voltages, being close to
zero, see also Figs.~\ref{Fig:2}(c) and \ref{Fig:3}(c). This
implies that the spin polarization of tunneling electrons is
significantly reduced. Such a behavior  is opposite to that in the
case of two strongly coupled dots, where TMR in the sequential
tunneling regime was found to be
considerable.~\cite{weymannPRB_DQDs}

\begin{figure}[t]
  \includegraphics[width=0.8\columnwidth]{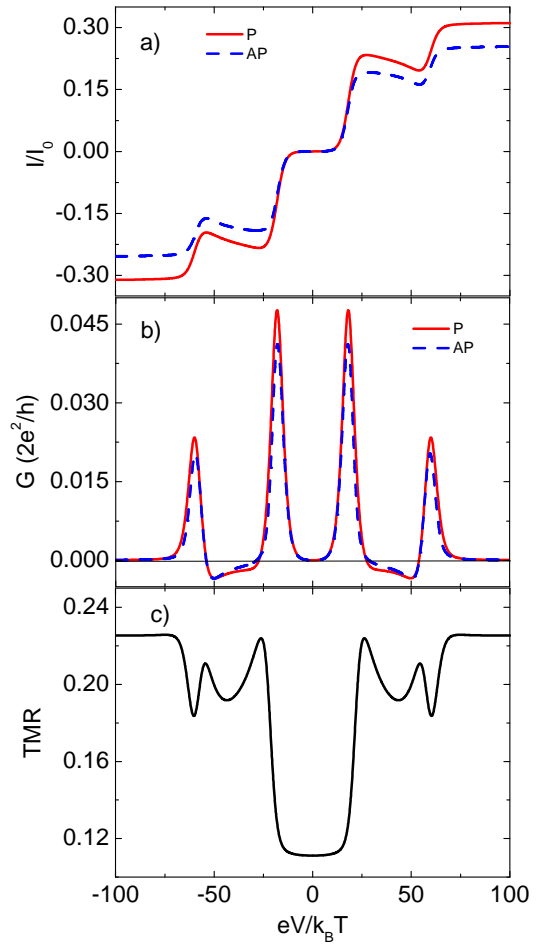}
  \caption{\label{Fig:7}
  (Color online) Current (a) and differential conductance (b)
  in the parallel (solid line) and antiparallel (dashed line)
  magnetic configurations, and tunnel
  magnetoresistance (c), calculated  as a function of the bias voltage
for $\alpha=0.25$.
  The other parameters are the same as in Fig.~\ref{Fig:6}.}
\end{figure}

\subsection{Double dots connected in parallel, $\alpha =1$}

In this subsection we consider the case when the dots are
connected in parallel, $\alpha=1$, see Fig.~\ref{Fig:1}. Virtual
tunneling processes between the two quantum dots through the leads
are now allowed. As mentioned earlier,  the off-diagonal matrix
elements of $\mathbf{\Gamma}_\beta^\sigma$ may be significantly
reduced due to suppression/cancellation of different
contributions, and hence $q<1$ in general. In the following we
assume $q=0.25$. However, we will also analyze how the transport
characteristics depend on the parameter $q$. For a given $q$ we
will examine two cases: symmetric case when all dot-lead couplings
are the same for given spin, i.e., $\alpha=1$, and asymmetric case
when there is a difference in the coupling of a given electrode to
the two dots, $\alpha\neq 0,1$. The latter case, referred to as
the intermediate geometry,  will be analyzed in the next
subsection.

The current, differential conductance and TMR for the case when
the DQD is empty at equilibrium, $E>0$, are shown in
Fig.~\ref{Fig:6}. The current exhibits a typical staircase-like
behavior [Fig.~\ref{Fig:6}(a)], which is also reflected in
well-defined peaks in the differential conductance located at the
positions $|eV|=2E$ and $|eV|=2E+2U_0$, see Fig.~\ref{Fig:6}(b).
The TMR is positive in the whole bias range and takes well-defined
values corresponding to different steps in the current-voltage
characteristics, see Fig.~\ref{Fig:6}(c). In the case of DQD
coupled in parallel, the TMR is generally larger than in the case
of serial connection discussed in the previous subsection,
especially for large voltages. This is due to a different role of
interdot hopping in these two geometries. As the hopping term is
crucial for transport in serial geometry, it  plays a less
important role in the parallel geometry. Since the hopping
parameter is independent of spin orientation, it leads to a
reduction of TMR in the serial geometry in comparison to that in
the parallel one, especially at large voltages.

\subsection{Double dots in a general (intermediate) geometry}

The situation changes considerably when the dot-lead couplings are
different, $\alpha\neq 1$, see Fig.~\ref{Fig:1}.
Figure~\ref{Fig:7} presents the current, differential conductance
and TMR as a function of bias voltage, calculated for $\alpha =
0.25$. As one can note,  the shape of curves describing current
and differential conductance reveals features  obtained above for
dots connected in series, see Fig.~\ref{Fig:2}, i.e. the negative
differential conductance and associated non-typical Coulomb steps
in current. However, the bias dependence of TMR is now different
and is more similar to that obtained for parallel geometry, except
for voltages $2E<|eV|<2E+2U_0$, i.e. between the resonances, where
the behavior of TMR is more complex and where one finds some
oscillations in TMR. Furthermore, for the intermediate geometry,
the TMR for larger bias voltages is quite significant, contrary to
the case of serial geometry, see Figs.~\ref{Fig:2}(c) and
\ref{Fig:3}(c), and is only a little  smaller than that in the
parallel geometry, see Fig.~\ref{Fig:6}.

\begin{figure}[t]
  \includegraphics[width=0.8\columnwidth]{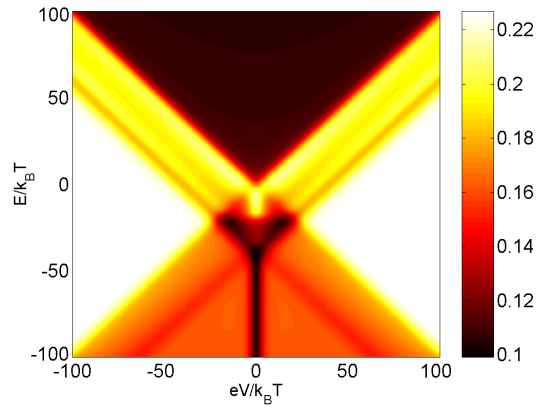}
  \caption{\label{Fig:8}
  (Color online) Tunnel magnetoresistance
  as a function of the bias voltage and the average level
  position, calculated for $\alpha=0.25$.
  The other parameters are the same as in Fig.~\ref{Fig:7}.}
\end{figure}

For completeness, in Fig.~\ref{Fig:8} we show the bias and gate
voltage dependence of the TMR. One can see that TMR displays well
defined structures consisting of regions where it is roughly
constant. In these transport regions the corresponding
current-voltage curves display plateaus, whereas at resonances the
corresponding TMR changes considerably.

\begin{figure}[t]
  \includegraphics[width=0.8\columnwidth]{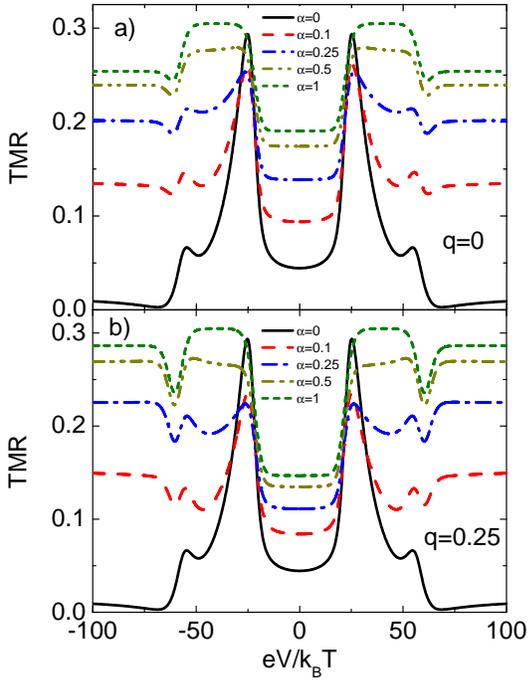}
  \caption{\label{Fig:9}
   (Color online) Bias voltage dependence of  TMR for indicated values of
   the asymmetry parameter $\alpha$ and for $q=0$ (a)
   and $q=0.25$ (b), calculated for $E=10k_BT$.
   The other parameters are the same as in Fig.~\ref{Fig:2}.}
\end{figure}

It is interesting to analyze how TMR depends on the asymmetry
factor $\alpha$. In Fig.~\ref{Fig:9} we present the bias
dependence of  TMR calculated for different values of $\alpha$ and
for the case when off-diagonal matrix elements $\Gamma_{\beta
12}^{\sigma}$ are zero ($q=0$) and finite ($q=0.25$), see
Fig.~\ref{Fig:9}(a) and (b), respectively. By changing $\alpha$
from $0$ to $1$, geometry of the system continuously changes from
serial to parallel one. Furthermore, with increasing $\alpha$ also
the magnitude of the off-diagonal matrix elements is increased,
see Eqs.~(\ref{Eq:coupling_left}) and (\ref{Eq:coupling_right}).
First of all, one can note that TMR generally increases with
increasing $\alpha$. This is associated with the fact that for
$\alpha \neq 0$ the electrons can tunnel between the leads through
just a single dot. Accordingly, the role of  interdot hopping,
which is independent of electron spin and therefore reduces TMR,
is diminished and TMR increases. In particular, when crossing over
from the serial to parallel geometry, the TMR at zero bias
increases, while sharp maxima in TMR at resonance voltages are
transformed into plateaus. This behavior can be observed in the
case  of $q=0$ and $q=0.25$, see Fig.~\ref{Fig:9}(a) and (b),
respectively. There are however some differences between these two
situations. The virtual tunneling processes between the two dots,
described by the nonzero parameter $q$, decrease the TMR at the
zero bias and increase it for large bias voltages, $|eV|>2E+2U_0$,
as compared to the case of $q=0$.

Another interesting feature visible in Fig.~\ref{Fig:9} is that
TMR for $\alpha=1$ at the side plateaus, i.e. for
$2E+2U_0>|eV|>2E$, has the same magnitude for both $q=0$ and
$q=0.25$ cases. This can be clearly shown by deriving an
approximate analytical formula for TMR in this voltage region. At
very low temperatures one can approximate the Fermi functions by
step functions and assume that the electrons tunnel only from one
side to the other.~\cite{weymannJPCM07} Then, one can show that
the TMR for $\alpha=1$ and $2E+2U_0>|eV|>2E$ is given by
\begin{equation}\label{Eq:TMR_analitical}
{\rm TMR}=\frac{8p^2}{5(1-p^2)} \;,
\end{equation}
which is exact at zero temperature. From the above formula follows
that TMR in the bias regime under consideration is independent of
the magnitude of the off-diagonal matrix elements. This is,
however, not true for the current, which for $\alpha=1$ is given
by
\begin{equation}
I^{\rm P}=\frac{e\Gamma}{\hbar}\frac{2+q}{5}\;,
\end{equation}
in the parallel and
\begin{equation}
I^{\rm AP}=\frac{e\Gamma}{\hbar}\frac{(2+q)(1-p^2)}{5+3p^2}\;,
\end{equation}
in the antiparallel magnetic configurations.

\begin{figure}[t]
  \includegraphics[width=0.9\columnwidth]{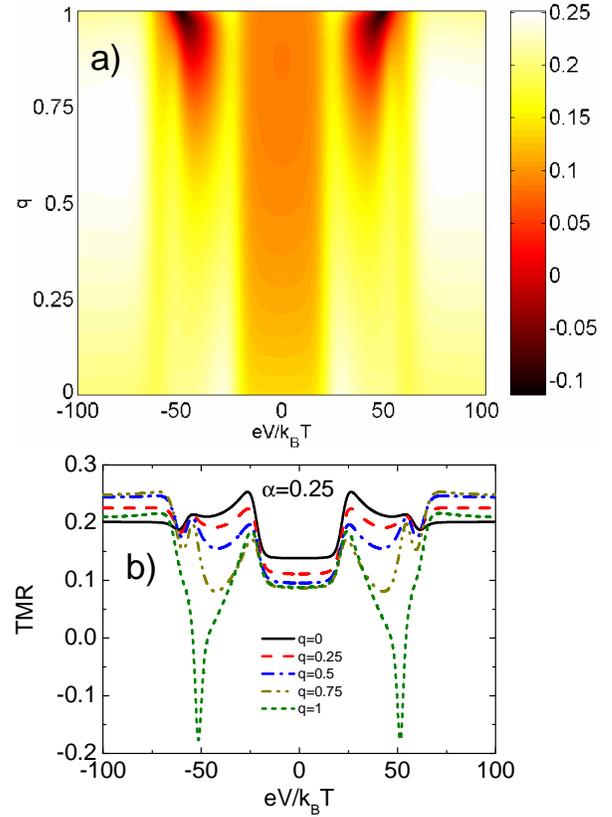}
  \caption{\label{Fig:10}
  (Color online) (a) Tunnel magnetoresistance as a
  function of the bias voltage and parameter $q$, calculated for
  for $\alpha=0.25$.
  (b) Cross-sections of TMR for several values of $q$
  as indicated in the figure.
  The other parameters are the same as in Fig.~\ref{Fig:6}.}
\end{figure}

The interference effects due to virtual first-order processes
between the two dots can significantly affect TMR, especially for
larger values of $q$. This is shown in Fig.~\ref{Fig:10}, which
depicts TMR as a function of the bias voltage and parameter $q$,
together with various cross-sections. It can be seen that with
increasing the magnitude of the off-diagonal matrix elements, the
TMR generally increases in the low bias voltage regime, $|eV|<2E$,
and for larger voltages, $|eV|>2E+2U_0$, although the latter
dependence is not monotonic. However, for bias voltages where
transport occurs through charge states with single electron on the
double dot, $2E+2U_0>|eV|>2E$, TMR becomes suppressed with raising
$q$, and for $q=1$ we find a negative TMR effect. The negative TMR
develops approximately at the resonance, $|eV|\approx 2E+2U_0$,
where two-particle states of the double dot system start taking
part in transport.

\begin{figure}[t]
  \includegraphics[width=0.8\columnwidth]{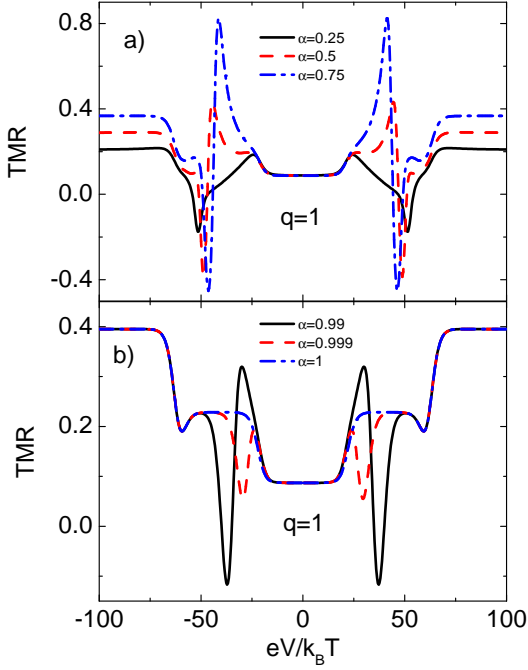}
  \caption{\label{Fig:11}
  (Color online) Bias voltage dependence of TMR
  calculated for $q=1$
  and for different values of the asymmetry parameter $\alpha$,
  as indicated in the figure.
  The other parameters are the same as in Fig.~\ref{Fig:2}.}
\end{figure}

Figure~\ref{Fig:10} was calculated for $\alpha=0.25$. However, it
turns our that the spin-dependent transport properties may also
strongly depend on the geometry of the system. This is especially
visible in TMR for maximum value of the parameter $q$, $q=1$. In
Fig.~\ref{Fig:11} we plot  TMR as a function of the bias voltage
for different values of $\alpha$ and for $q=1$. As one can note,
TMR is rather independent of $\alpha$ at low bias voltages, while
for larger bias, $|eV|>2E+2U_0$, TMR generally increases with
raising $\alpha$. This is however not the case for
$2E+2U_0>|eV|>2E$, where TMR strongly depends on the geometry of
the system. With crossing over from serial to parallel geometry,
the magnitude of negative TMR is increased and additional maximum
develops next to the resonance, $|eV|\approx 2E+2U_0$, which
transforms into plateau for $\alpha$ close to 1.

Interestingly, it can be also seen that tunnel magnetoresistance
is very sensitive to slight changes in the asymmetry parameter
$\alpha$, when the latter is close to unity. To understand this
behavior one needs to realize that the bare states of the two
quantum dots coupled directly (by the hopping term) or indirectly
(due to off-diagonal coupling matrix elements) hybridize in
molecular-like states. As a result, the bonding and anti-bonding
states emerge, the widths of which strongly depend on the dot-lead
coupling strengths and geometry of the system. In the case
considered here, the relative width of the bonding and
anti-bonding states varies with the parameter $\alpha$. When the
difference in the coupling of a given electrode to the two dots is
reduced, the width of the bonding state increases whereas that of
the anti-bonding state decreases. In particular, in the limit of
$\alpha=1$, the anti-bonding state becomes totally decoupled from
the leads, while the bonding state acquires width of the order of
$2\Gamma$. In this limit the above mentioned features of the TMR
disappear and tunnel magnetoresistance is constant for
$2E+2U_0>|eV|>2E$. In other words, the high sensitivity of the TMR
with respect to the system's geometry is associated with the fact
that the interference conditions for electron waves transmitted
through the two dots become modified when changing $\alpha$.


\section{Summary and conclusions}


We have analyzed the spin-polarized transport properties of double
quantum dots weakly coupled to each other and to external leads.
Using the real-time diagrammatic technique we have calculated the
conductance and tunnel magnetoresistance in the parallel, serial
and intermediate geometries of double quantum dots. Moreover, we
have taken into account the effects of virtual tunneling processes
between the two dots taking place through the states in the leads.
Such processes are absent in serial geometry and become maximum
for parallel geometry.

In the case of double quantum dots coupled in series we have found
a negative tunnel magnetoresistance at the resonance and negative
differential conductance for transport voltages where
single-particle double-dot states take part in transport. On the
other hand, for parallel geometry of the system, both the negative
TMR and negative differential conductance vanish. The above
effects may be restored in an intermediate geometry and strongly
depend on the magnitude of the virtual processes between the two
dots. Furthermore, in the case when virtual processes are maximal,
we have found a strong dependence of the TMR on the geometry of
the system, especially for geometries very close to the parallel
one.


\begin{acknowledgements}
This work, as part of the European Science Foundation EUROCORES
Programme SPINTRA, was supported by funds from the Ministry of
Science and Higher Education as a research project in years
2006-2009 and the EC Sixth Framework Programme, under Contract N.
ERAS-CT-2003-980409. P.T. also acknowledges support by funds from
the Ministry of Science and Higher Education as a research project
in years 2009-2011. I.W. acknowledges support from the Alexander
von Humboldt Foundation, the Foundation for Polish Science and the
Ministry of Science and Higher Education through research project
N N202 169536 in years 2008-2010. Financial support by the
Excellence Cluster "Nanosystems Initiative Munich (NIM)" is
gratefully acknowledged.
\end{acknowledgements}


\appendix

\section{Examples of the first-order self energies}

\begin{figure}[t]
  \includegraphics[width=0.84\columnwidth]{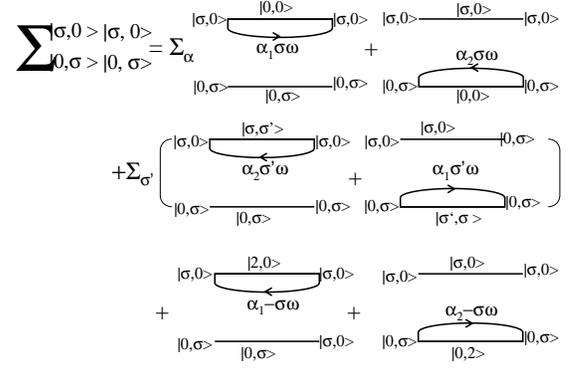}
  \caption{\label{Fig:12} Graphical representation of the self-energy
  $\Sigma^{|\sigma,0\rangle|\sigma,0\rangle}_{|0,\sigma\rangle|0,\sigma\rangle}$ (a).
  The summations are over lead and spin degrees of freedom $\alpha=L,R$
  and $\sigma'=\uparrow,\downarrow$. Each tunneling line carries lead index
  $\alpha$, spin $\sigma$ and frequency $\omega$.}
\end{figure}

\begin{figure}[t]
  \includegraphics[width=0.84\columnwidth]{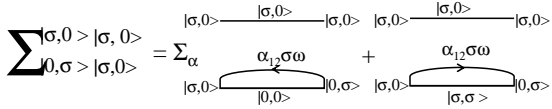}
  \caption{\label{Fig:13} Graphical equation for the self-energy
  $\Sigma^{|\sigma,0\rangle|\sigma,0\rangle}_{|0,\sigma\rangle|\sigma,0\rangle}$.}
\end{figure}

In order to analyze the transport properties one needs to
calculate the respective self-energies using corresponding
diagrammatic rules.~\cite{diagrams,weymannPRB05} In the sequential
tunneling regime only the first-order self-energies determine the
transport characteristics. Here, we present explicitly two
examples of first-order self-energies. Generally, the
self-energies are complex -- their imaginary part may be related
to transition rates, whereas the real part may be associated with
with various renormalization effects. The graphical representation
of the self-energy
$\Sigma^{|\sigma,0\rangle|\sigma,0\rangle}_{|0,\sigma\rangle|0,\sigma\rangle}$
is displayed in Fig.~\ref{Fig:12}, while analytically it is given
by
\begin{widetext}
\begin{eqnarray}\label{diagram1}
\Sigma^{|\sigma,0\rangle|\sigma,0\rangle}_
    {|0,\sigma\rangle|0,\sigma\rangle} &=& \sum_{\alpha,\sigma'}
    \left\{
    P^{-,\sigma}_{\alpha_{22}}({\varepsilon_{1\sigma}})
    -P^{-,\sigma}_{\alpha_{11}}(\varepsilon_{2\sigma})
   +P^{+,\sigma'}_{\alpha_{22}}({\varepsilon_{1\sigma}+
    \varepsilon_{2\sigma'}+U_0-\varepsilon_{2\sigma}})
    -P^{+,\sigma'}_{\alpha_{11}}(\varepsilon_{1\sigma'}+
    \varepsilon_{2\sigma}+U_0-\varepsilon_{1\sigma})
 \right.
\nonumber \\
    &&+P^{+,\bar{\sigma}}_{\alpha_{11}}(\varepsilon_{1\sigma}+
    \varepsilon_{1\bar{\sigma}}+U_1-\varepsilon_{2\sigma})
    -P^{+,\bar{\sigma}}_{\alpha_{22}}
    (\varepsilon_{2\sigma}+
    \varepsilon_{2\bar{\sigma}}+U_2-\varepsilon_{1\sigma})\nonumber\\
    &&-i\pi\left[\gamma^{-,\sigma}_{\alpha_{22}}(\varepsilon_{1\sigma})
    +\gamma^{-,\sigma}_{\alpha_{11}}(\varepsilon_{2\sigma})
    +\gamma^{+,\sigma'}_{\alpha_{22}}(\varepsilon_{1\sigma}+
    \varepsilon_{2\sigma'}+U_0-\varepsilon_{2\sigma})
    +\gamma^{+,\sigma'}_{\alpha_{11}}(\varepsilon_{1\sigma'}+
    \varepsilon_{2\sigma}+U_0-\varepsilon_{1\sigma})
\nonumber\right.\\
    &&\left.\left.
    +\gamma^{+,\bar{\sigma}}_{\alpha_{11}}
    (\varepsilon_{1\sigma}+
    \varepsilon_{1\bar{\sigma}}+U_1-\varepsilon_{2\sigma})
    +\gamma^{+,\bar{\sigma}}_{\alpha_{22}}
    (\varepsilon_{2\sigma}+
    \varepsilon_{2\bar{\sigma}}+U_2-\varepsilon_{1\sigma})\right]\right\} \,.
\end{eqnarray}
\end{widetext}
In the above expression the states $|2,0\rangle$, $|0,2\rangle$
denote doubly occupied first and second dot, respectively, and,
because of large intradot Coulomb repulsion, are only considered
as virtual ones (intermediate states). In Fig.~\ref{Fig:13} we
also show an example of self-energy including off-diagonal matrix
elements, $\mathbf{\Gamma}_\beta^\sigma$,
$\Sigma^{|\sigma,0\rangle|\sigma,0\rangle}_{|0,\sigma\rangle|\sigma,0\rangle}$,
it is given by
\begin{eqnarray}\label{diagram2}
    \Sigma^{|\sigma,0\rangle|\sigma,0\rangle}_
    {|0,\sigma\rangle|\sigma,0\rangle}=\sum_{\alpha}
    \left\{P^{-,\sigma}_{\alpha_{12}}
    ({\varepsilon_{1\sigma}})
    -  i\pi\gamma^{-,\sigma}_{\alpha_{12}}(\varepsilon_{1\sigma})\right.
\nonumber \\
       +\left.P^{+,\sigma}_{\alpha_{12}}
    ({\varepsilon_{2\sigma}+U_0})
    +i\pi\gamma^{+,\sigma}_{\alpha_{12}}(\varepsilon_{2\sigma}+U_0)\right\}\,,
\end{eqnarray}
with
\begin{equation}
   P^{\pm,\sigma}_{\alpha_{ij}}({x})=\pm\frac{\Gamma_{\alpha
   ij}^{\sigma}}{2\pi}
   \left[{\rm Re}\Psi\left(\frac{1}{2}+i\frac{x-\mu_\alpha}{2\pi
   k_BT}\right)-\log\left({\frac{D}{2\pi k_BT}}\right)\right],
\end{equation}
\begin{eqnarray}
\gamma^{\pm,\sigma}_{\alpha_{ij}}(x)=\frac{\Gamma_{\alpha
ij}^{\sigma}}{2\pi}f^\pm\left(x-\mu_\alpha\right),
\end{eqnarray}
where $f^+(x)$ stands for Fermi-Dirac distribution,
$f^-(x)=1-f^+(x)$, and $\Psi(x)$ denotes the digamma function.
Here, $D$ is the cut-off parameter which can be identified with
on-level Coulomb repulsion $U$. It is worth to note that the
self-energies do not depend on the cut-off parameter because the
terms depending on $D$ cancel in pairs, see Eqs.~(\ref{diagram1})
and (\ref{diagram2}).


\end{document}